# ARE ENTROPIC BELL INEQUALITIES IMPLIED BY THE SECOND LAW?

*Ian T. Durham[†]*
[†]*Department of Physics*
*Saint Anselm College*
*100 Saint Anselm Drive, Box 1759*
*Manchester, NH 03102*
*idurham@anselm.edu*
(Updated: March 31, 2007)

**Abstract**

Bell's inequalities, in the form given by Cerf and Adami, are derived from the combination of the second law of thermodynamics and the Markov postulate. Violations of these inequalities are discussed in terms of the mixing characteristics of the density operators. A quantum mechanical bound for a system of three qubits, two of which are entangled, is found and is shown to lead to strong subadditivity in one case. Finally, in interpreting the results, the role the uncertainty relations play in the definition of entropy and the second law is discussed, demonstrating a link to Bell's inequalities. Additionally the second law is shown to be dependent on the type of measurement in quantum systems. Specifically, a more general statement of the second law appropriate to both quantum and classical systems is suggested.

## I. BACKGROUND

In this paper I: a.) establish that the second law depends on the type of measurement when applied to quantum systems; b.) note that the most general statement of the second law that applies to both quantum and classical systems is the requirement that the entropy of mixing never decreases, assuming the quantum measurements are projective; c.) show that in quantum systems this follows from the specific nature of the density operators; d.) determine that the Cerf-Adami inequalities follow from the traditional *classical* version of the second law that *includes* the positivity condition and thus the Markov postulate; e.) show that this classical version of the second law is partly a consequence of the uncertainty relation since this leads to the positivity condition, thus demonstrating a relation between these inequalities and the uncertainty relation; f.) determine a quantum mechanical bound for the Cerf-Adami inequalities for a quantum system of three qubits, two of which are entangled; and g.) show that with this bound one can arrive at a statement of strong subadditivity.

It has been argued that the laws governing entanglement may well be thermodynamic in nature, or, at the very least, possess thermodynamic corollaries [1]. In addition, Bell's inequalities have been derived from considerations of the spin-statistics theorem by O'Hara [2] which, of course, is at the heart of quantum statistical mechanics. Classically, Landauer's principle as an analogue to the second law of thermodynamics (law of entropy increase [3]) has been recently defended by Bennett [4]. It seems natural, then, that a fuller articulation of the thermodynamic nature of entanglement would be found in quantum information theory. On the other hand, the role that the second law of thermodynamics plays in its more traditional entropic sense is still unclear. Part of the problem stems from misunderstandings and misrepresentations of the second law. For instance, from a pedagogical standpoint it is often the case that the second law is *implied* (though perhaps never explicitly stated) as being a fundamental law [5] when, in fact, it is actually just a strong statement on the behavior of probabilities [6]. When viewed in this statistical manner, entropy is interpreted as simply being another way of measuring the number of configurations of a system or, rather, the probabilities that it will be *in* one of those configurations (whether those probabilities are Bayesian or frequentist is immaterial in this case). The fundamental assumption of statistical mechanics implies that all accessible states of a uniform probability distribution are equally likely in the long run. If we take this to be the case and work solely with such distributions, we can represent the number of configurations of a system by the multiplicity, $\Omega$, and the entropy can be represented as

$$S = k \ln \Omega \qquad (1)$$

often called the Boltzmann entropy, which is entirely equivalent[1] to the form

$$S = -k \sum_i p(x) \ln p(x). \qquad (2)$$

where the sum is only over these accessible states. Equation (2) is sometimes referred to as the Boltzmann-Gibbs-Shannon entropy since its basic form encompasses the definitions developed by Boltzmann [7], Gibbs [8], and Shannon [9]. There are, of course, numerous other interpretations of and definitions for entropy. Since this is not intended as an analysis of entropy itself, I will only mention two of these. In particular I reference the work of Penrose [10] whose interpretation utilizes the basic form encompassed in equation (2). Also, by taking the forms of Boltzmann, Gibbs, and Shannon as functionally similar enough for the purposes outlined here I am deliberately *not* addressing Carnap's argument [11] that a fine-grained analysis of these definitions find them to be inconsistent in certain circumstances. I do, however, point out the fact that inconsistencies can arise in these classical definitions. It is simply immaterial for the present argument since my goal is rather to contrast the quantum with the classical in an analysis of the second law.

    Let us first discuss the way in which the second law of thermodynamics may be discussed in terms of this definition for entropy. As Landau and Lifshitz point out, there are inherent difficulties in the interpretation of equation (1) in terms of the units (i.e. the units are entirely wrapped up in the multiplicative constant, $k$). As such, the only uniquely determined quantities that do not depend on the choice of units are *differences* in entropy, i.e. the changes in entropy brought about by some process [3]. Rather than simply equate the second law with a non-negative entropy, a stronger formulation is to define it in terms of the *entropy of mixing* which is the entropy created in the process of mixing two systems. Consider, for example, two systems *A* and *B* with classical

---

[1] See [6]. For an isolated system only the accessible states have non-zero probability and so the sum in equation (2) can be restricted to these accessible states giving

$$S = -k \sum_s \frac{1}{\Omega} \ln \frac{1}{\Omega} = \frac{k \ln \Omega}{\Omega} \sum_s (1).$$

But the number of terms is just $\Omega$ and thus equation (2) reduces to equation (1).

thermodynamic entropies $S(A)$ and $S(B)$ that are initially separated and then allowed to interact in some manner (for example, two ideal gases separated by a barrier that is later removed). Speaking in purely statistical terms, once the barrier is removed and the two systems mix, the total number of configurations for the combined system must either be equal to or greater than the total number of configurations of the two separate systems. Since the entropy is just another way of representing the number of configurations, we can refer to this *difference* in entropy as the *entropy of mixing*[2]. We can write the second law, then, in terms of the entropy of mixing as

$$S_{mix} = \Delta S_{total} \geq 0 \qquad (3)$$

where the equality holds if systems *A* and *B* are identical. For example, then, if the systems represent ideal gases and entropy is a method for expressing the probability that a system will be in a given state, the individual entropies provide a method for expressing the probability distributions of the two systems. If the two gases were the same species and otherwise identical prior to mixing, there would be no difference between the two probability distributions and thus no entropy of mixing (technically the multiplicity increases slightly but the factor is negligible and thus it is approximately zero, but always positive regardless).

From equations (1) and (2) numerous approximate expressions for the entropy of complicated systems have been derived. One of the best known is the Sackur-Tetrode equation that gives the entropy of an ideal gas. However, as with most of these approximations, it has limited applicability demonstrated by the fact that it is negative for temperatures below roughly 0.01 K [6]. This is possible, despite the positive definite nature of equation (2), because the Sackur-Tetrode equation is semi-empirical. I discuss negative classical entropies further in the final section. Nonetheless, these approximations often lead to erroneous conclusions concerning the possibility of violating equation (3) by producing a negative entropy in the first term of the difference. As a note, there are two different aspects of the second law wrapped up in equation (3). The first is that entropy changes tend to be positive (entropy tends to increase). The second is that the entropy itself, regardless of its change, is a positive value. Though no such situation has ever been observed, one might envision an alternate universe in which the entropies are always positive and yet equation (3) is violated. Otherwise, no violation of this inequality has ever been observed macroscopically and it is henceforth taken as a representation of the second law. Later I will discuss a stronger statement of the second law that bridges the gap between quantum and classical domains and the fact that the Markov postulate is an essential requirement for statistical entropy to obey the second law.

## II. ENTROPY AND THE SECOND LAW

In information theory it is usual to represent entropy in the binary sense as articulated by Shannon [9],

---

[2] See, for example, [6], p. 79.

$$H(X) = -\sum_x p(x) \log p(x) \tag{4}$$

where the logarithm is taken to be base-two. We define the relative entropy [12]

$$H(p(x,y) \| p(x)p(y)) \equiv -\sum_{x,y} p(x,y) \log \frac{p(x,y)}{p(x)p(y)} \tag{5}$$
$$= H(p(x)) + H(p(y)) - H(p(x,y))$$

to be a measure of the "offset" of the probability distribution over two indices, $x$ and $y$, from the probability distributions of the individual indices themselves. As in [12], we define $-0\log 0 \equiv 0$ and $-p(x,y)\log 0 \equiv +\infty$ if $p(x,y) > 0$. Since this represents an offset of the probability distributions it is zero when these distributions are the same. As such, it is simply a binary form of the entropy of mixing and thus obeys equation (3) which I will refer to as the *positivity condition* and have already taken as representative of the second law and noted is well established [3, 10].

The relative entropy can be expressed in a number of ways including as the *mutual* entropy which represents the mutual information of two systems. As such, consider two systems, $A$ and $B$, that are measured on indices $x$ and $y$ respectively. We can define the mutual entropy then [12] as

$$H(X:Y) \equiv H(X) + H(Y) - H(X,Y) \tag{6}$$

where

$$H(X,Y) \equiv -\sum_{x,y} p(x,y) \log p(x,y) . \tag{7}$$

The positivity condition in combination with equations (4), (5), and (6), implies that

$$H(X:Y) = H(p(x,y) \| p(x)p(y)) \geq 0 \tag{8}$$

where the equality holds only when $X$ and $Y$ are taken to be independent random variables *or*, if when $X = Y$, we take $p(x,y) = 0$.

Consider now *three* systems that I will call $X$, $Y$, and $Z$, each having a corresponding entropy $H(X)$, $H(Y)$, and $H(Z)$. We can therefore write mutual entropies $H(Y:Z)$ and $H(X:Z)$ in addition to $H(X:Y)$. If $X \Rightarrow Y \Rightarrow Z$ is a Markov chain (such that $Y$ is only accessible from $X$ and $Z$ is only accessible from $Y$) then so is $Z \Rightarrow Y \Rightarrow X$ (where the reverse is true) and it follows that

$$H(Z:Y) \geq H(Z:X) \tag{9}$$

which is sometimes known as the data pipelining inequality which, given the property that $H(X:Y) = H(Y:X)$ [12], can be written $H(Y:Z) \geq H(X:Z)$. Likewise the Markov chain assumption also implies $H(X:Y) \geq H(X:Z)$. Trivially, then, we can write

$$H(X:Y) + H(Y:Z) \geq H(X:Z). \tag{10}$$

Equation (10) also follows trivially from the combination of the data pipelining inequality and the positivity condition which I have taken to represent the second law.

Another basic property of Shannon entropies is that $H(X:Y) \leq H(Y)$ and $H(Y:Z) \leq H(Y)$. Combining this with equation (10) we can write

$$H(X:Y) + H(Y:Z) - H(X:Z) \leq H(Y). \tag{11}$$

Likewise, given that $H(X:Y) \leq H(X)$, $H(X:Z) \leq H(X)$, $H(X:Z) \leq H(Z)$, and $H(Y:Z) \leq H(Z)$, we can write

$$H(X:Y) + H(X:Z) - H(Y:Z) \leq H(X) \tag{12}$$

$$-H(X:Y) + H(X:Z) + H(Y:Z) \leq H(Z). \tag{13}$$

If $X$, $Y$, and $Z$ represent uniform distributions, then $H(X) = H(Y) = H(Z) = 1$ and we arrive at the inequality found by Cerf and Adami (in much the same manner) [13],

$$|H(X:Y) - H(X:Z)| + H(Y:Z) \leq 1 \tag{14}$$

and which they have shown to be in perfect analogy to the usual form of Bell's inequalities. In addition, these greatly resemble the Braunstein-Caves inequalities [14]. These inequalities have been directly derived from the basic definition of entropy and three initial conditions: the positivity condition (the second law) and the Markov postulate. However, as Penrose rigorously points out, any non-decreasing statistical entropy must satisfy two additional constraints: the Markov chain must be *deterministic*, and only the *number* of individual systems can be observed and *not* their identities [10]. Note the latter two conditions really establish the *positivity condition* and are thus prior to the *definition* of entropy. Ultimately, the relational order is thus:

$$M^{det} + I + \text{def}(S) \Rightarrow S_{stat} \geq 0 \Rightarrow \text{Bell's inequalities} \tag{15}$$

where $M^{det}$ is a *deterministic Markov chain*, $I$ represents indistinguishability, and def($S$) represents the basic definition of statistical entropy (which in this case is the Shannon entropy). Bell's inequalities follow directly from the non-negativity of the statistical entropy and thus the second law, with their deterministic nature stemming from the constraint on the Markov chain (required for non-negativity).

### III. QUANTUM – CLASSICAL CONTRAST

We can make the probabilities in equations (2) and (4) the diagonal components of a matrix, holding all non-diagonal components zero for the moment. Equation (2), then becomes

$$S(\rho_C) = -\sum_x \rho_C(x)\ln\rho_C(x). \tag{16}$$

With this definition we can construct a stronger positivity argument (even though it appears obvious that this cannot ever be negative). First note that given any POVM, $A$, defined on the Borel subsets of a space, $X$, a classical probability density $\rho_C$ can be determined from a quantum density operator, $\rho$, as

$$\rho_C = \text{tr}(\rho A). \tag{17}$$

This, in conjunction with the usually defined expectation value for the POVM, makes the classical entropy semi-bounded such that

$$S(\rho_C) \geq 0. \tag{18}$$

This theorem is proven by Schroeck [15] and constitutes a stronger version of the positivity condition that is at the heart of the second law. In addition, with equation (17), we have constructed a link between the classical entropies as utilized in the Cerf-Adami inequalities and the quantum entropies. Specifically, equation (17) allows us to write the von Neumann (quantum) entropy as

$$S = -\text{tr}(\rho \ln \rho) \tag{19}$$

where the density operators are for the quantum system here. If these density operators can be diagonalized (i.e. if they are separable), they reduce to the classical probabilities and the trace represents the sum. In this way equation (19) can reduce to equation (16) in such an instance.

To further extend this let us consider this in light of the entropy of mixing as described earlier. Shannon [9] defines the information (or information gain) of a measurement as the change in entropy which is analogous to equation (3). The question becomes: *what* about entropy imbues it with its properties? In particular, if the entropy increases with mixing, what is the corresponding behavior of the probabilities or, more generally, the density operator? As it turns out the density operator itself satisfies a positivity condition as well as a trace condition. In particular, an operator $\rho$ is the density operator associated with some ensemble if and only if it has a trace equal to one (the trace condition) and is a positive operator (the positivity condition). Conversely if an operator $\rho$ satisfies the trace and positivity conditions it has a spectral decomposition

$$\rho = \sum_j \lambda_j |j\rangle\langle j| \tag{20}$$

where $|j\rangle$ are orthogonal and $\lambda_j$ are the real, non-negative eigenvalues of $\rho$ that satisfy $\sum_j \lambda_j = 1$. A full proof is found in [12]. Note, however, that if we do not necessarily sum over all *j* but sum only up to a value *N*, it is possible to arrive at Ruch's [16] definition of the relation ≺

$$\rho \prec \rho' \text{ iff } \sum_{i=1}^{N} \lambda_i \geq \sum_{i=1}^{N} \lambda'_i \qquad (21)$$

where $\rho \prec \rho'$ reads "$\rho'$ is more mixed than $\rho$." To fully clarify what the inequality involving the summations means, consider this example: let *r* have eigenvalues 1/2, 1/4, 1/4, 0, 0, … and *p* have eigenvalues 2/3, 1/6, 1/6, 0, 0, … Notice that in each case the sum of all the eigenvalues is 1 and so both satisfy the trace condition. However, notice that the first eigenvalue for *p* is greater than the first eigenvalue for *r*. Likewise, the sum of the first *two* eigenvalues for *p* is still greater than the sum of the first two for *r*. Only with the sum including the third eigenvalue do we arrive at the equality. In this case, based on equation (21), we would say that *r* is more mixed than *p* based on the *ordering* of their eigenvalues [15].

As it turns out, $\rho \prec \rho'$ can imply under certain circumstances that some function $f(\rho) \leq f(\rho')$ and thus these functions are said to be mixing homomorphic (that is order preserving). Since the von Neumann entropy is mixing homomorphic [15] it is a measure of the order or disorder of a system, which is consistent with at least one interpretation of generalized entropies. In fact, this fact can be used to show that the von Neumann entropy itself is invariant under unitary evolution [15], though the density operator itself may not be. In fact in the case of projective measurements the density operator evolves unitarily and the entropy increases [12]:

$$S(\rho') \geq S(\rho). \qquad (22)$$

Thus projective measurements have the same effect on the entropy as mixing and equation (22) can be interpreted as a generalization of equation (3) which is, again, taken to be a statement of the second law. These definitions will now allow us to analyze violations of equation (14).

Let us now take equation (7) to define the *joint entropy* of a pair of random variables. When combined with equation (6), equation (14) can be rewritten as

$$|H(Y) - H(Z) + H(X,Z) - H(X,Y)| + H(Y) + H(Z) - H(Y,Z) \leq 1. \qquad (23)$$

Now define the von Neumann entropy of the joint state of two qubits *A* and *B* to be $S(\rho^{AB}) = S(A,B) = -tr(\rho^{AB} \log \rho^{AB})$ where the logarithm in this case is base two. Let us then naïvely assume we can use quantum entropies in equation (23) and consider qubits *A*, *B*, and *C*. If we then consider a state in which two of the qubits, *B* and *C*, are in an entangled state $(|00\rangle + |11\rangle)/\sqrt{2}$, the joint quantum entropy, *S(B,C)* = 0. However, the individual density operators for *B* and *C* are *I*/2 and thus their individual entropies, *S(B)* = *S(C)* = 1. Since everything else on the left side is contained within the absolute value, this inequality is violated in this instance with a minimum value of 2 on the left side. To clarify, for a system of three qubits, *A*, *B*, and *C*, with *B* and *C* being entangled (while *A* is not entangled with anything), the *quantum mechanical prediction* for equation (14) is

$$|S(A:B) - S(A:C)| + S(B:C) \leq 2. \tag{24}$$

This is similar to the contrast between the quantum and classical predictions of the usual Bell inequalities, though in that case the difference is an order of $\sqrt{2}$.

It is, in fact, possible to obtain strong subadditivity from this. Take the following specific example. Just as equation (14) is a generalization of equation (12), equation (24) is a generalization of inequalities including one similar to equation (12)

$$S(A:B) + S(A:C) - S(B:C) \leq 2. \tag{25}$$

Since the mutual entropies are all numbers between 0 and 1, the first two terms satisfy the inequality by themselves (i.e. regardless of the value for $S(B:C)$). If system $A$ is in a state such that its entropy is one, it follows that

$$S(A:B) + S(A:C) \leq 2S(A) \tag{26}$$

which can be taken as a statement of strong subadditivity [12].

## IV. INTERPRETING VIOLATIONS

Clearly the violation of the Cerf-Adami inequalities by quantum systems is a direct consequence of the entangled and mixing characteristics of the density operators involved since the von Neumann entropies are mixing homomorphic. In fact the quantum entropy only increases for *projective* measurements. It can *decrease* for generalized measurements. As such, projective measurements are equivalent to an increase in mixing. Another way of saying this is to note that quantum *conditional* entropies can be *negative*. In relation to the second law, then, quantum systems approximate classical behavior in the case of projective measurements (this does *not* mean projective measurements are in any way classical, simply that they are consistent with the positivity condition inherent in the second law!). Nonetheless, since generalized measurements aren't remotely like this, it appears that the definition of the second law based on the non-decrease of entropy *is measurement dependent when applied to quantum systems*. In addition, since quantum systems allow for the existence of negative entropies, not all quantum systems obey the Markov postulate which is a requirement for a non-decreasing entropy. As such some quantum systems are not deterministic (local), a fact that is obviously well-known. Nonetheless, the entropy still increases for systems whose mixing increases. As such, *this is the more general statement of the second law that applies equally well to both quantum and classical systems provided the quantum measurements are projective*.

As a note, however, the positivity of classical entropies actually has its root in quantum mechanics. The multiplicity of a system as utilized in the definition of entropy given in equation (1) can be written in terms of the volume of the phase space [3, 6]

$$S = k \ln \frac{(\Delta p \Delta q)^d}{h^d} \tag{27}$$

where *d* is the number of degrees of freedom for the system. The quantities $\Delta p$ and $\Delta q$ are the spread in momentum and position, respectively, of the wavefunction describing a particle or particles in the system [6]. In and of itself, the spread is a simple classical statistical quantity. However, since we are describing this in terms of a quantum mechanical wavefunction, the product of these two spreads exhibits quantum behavior in the form of the uncertainty principle. In other words, if the position and momentum are operators they satisfy the uncertainty relation. A deeper analysis of the various mathematical interpretations of the spread and its relation to uncertainty can be found in [15]. Generally, the most consistent interpretation of the spread appears to be one in terms of variance. In any case, the uncertainty relation provides a lower bound of 1 for the argument of the logarithm in equation (27), thereby establishing the positivity condition for entropy. In fact, Landau and Lifshitz have suggested that the second law *requires* some inequality in terms of *h* [3] and subsequent authors have suggested the uncertainty relation is simply the quantum mechanical statement of the second law. In any case, since the positivity condition is inherent in the Cerf-Adami inequalities, it is clear that the uncertainty principle is also integral to their formulation. In addition, one manner in which the second law can be described is by the fact that the entropy of a subsystem is proportional to the *size* of the system as a whole. Korepin used this to derive the entanglement entropy from the second law [17] for one-dimensional gap-less models.

In summary, I have clarified the nature of the second law of thermodynamics in quantum systems with a particular emphasis on its nature in relation to entangled systems described by inequalities of the form derived by Cerf and Adami. Specifically, I have

i. established that the second law depends on the type of measurement when applied to quantum systems;
ii. noted that the most general statement of the second law that applies to both quantum and classical systems is the requirement that the entropy of mixing never decreases, assuming the quantum measurements are projective;
iii. shown that in quantum systems this follows from the specific nature of the density operators;
iv. determined that the Cerf-Adami inequalities follow from the traditional *classical* version of the second law that *includes* the positivity condition and thus the Markov postulate;
v. shown that this classical version of the second law is partly a consequence of the uncertainty relation since this leads to the positivity condition, thus demonstrating a relation between these inequalities and the uncertainty relation;
vi. determined the quantum mechanical bound for the Cerf-Adami inequalities for a quantum system of three qubits, two of which are entangled;
vii. shown that with this bound one can arrive at a statement of strong subadditivity;

I would like to thank Paul O'Hara, Frank Schroeck, Barry Sanders, and an anonymous reviewer of a related paper for the extensive comments that led to this more rigorous version of this paper. I additionally thank Barry for suggesting where this research might lead.